# Can we assess research using open scientific knowledge graphs? A case study within the Italian National Scientific Qualification

Federica Bologna, Department of Classical Philology and Italian Studies, University of Bologna, Bologna, Italy - email: federica.bologna3@studio.unibo.it - ORCID: 0000-0002-3845-8266

Angelo Di Iorio, Department of Computer Science and Engineering, University of Bologna, Bologna, Italy - email: angelo.diiorio@unibo.it - ORCID: 0000-0002-6893-7452

Silvio Peroni, Research Centre for Open Scholarly Metadata, Department of Classical Philology and Italian Studies, University of Bologna, Bologna, Italy - email: silvio.peroni@unibo.it - ORCID: 0000-0003-0530-4305

Francesco Poggi, Department of Communication and Economics, University of Modena and Reggio Emilia, Italy - email: francesco.poggi@unimore.it - ORCID: 0000-0001-6577-5606

## Abstract

The need for open scientific knowledge graphs is ever increasing. While there are large repositories of open access articles and free publication indexes, there are still few free knowledge graphs exposing citation networks, and often their coverage is partial. Consequently, most evaluation processes based on citation counts rely on commercial citation databases. Things are changing thanks to the Initiative for Open Citations (I4OC, https://i4oc.org) and the Initiative for Open Abstracts (I4OA, https://i4oa.org), whose goal is to campaign for scholarly publishers to open the reference lists and the other metadata of their articles. This paper investigates the growth of the open bibliographic metadata and open citations in two scientific knowledge graphs, OpenCitations' COCI and Crossref, with an experiment on the Italian National Scientific Qualification (NSQ), the National process for University Professor qualification which uses data from commercial indexes. We simulated the procedure by only using such open data and explored similarities and differences with the official results. The outcomes of the experiment show that the amount of open bibliographic metadata and open citation data currently available in the two scientific knowledge graphs adopted is not yet enough for obtaining results similar to those provided using commercial databases.

## Acknowledgements

This work has been supported by the University fund for Research 2020 (FAR) of the University of Modena and Reggio Emilia, and by the European Union's Horizon 2020 research and innovation program under grant agreement No. 101017452.

# Introduction

In recent years, citation indexes have become increasingly important to evaluate the scientific performance of institutions. Bibliometrics such as citation count and h-index are being used to assess individual scholars and research bodies when allocating funding at the national level. For instance, in Germany, performance-based funding systems take into consideration the impact factor of the publications. In Finland, bibliometrics and citation indexes are being used in the reallocation system. In Norway, a two-level bibliometrics is employed for similar purposes (Vieira et al., 2014).

Several studies have analyzed the connection between citation indexes and research assessment. For example, some have focused on bibliometrics in relation to the results of the Research Assessment Exercise (RAE) in Britain (Norris & Oppenheim, 2003; Taylor, 2011), the Italian Triennial Assessment Exercise (VTR) (Abramo et al., 2009; Franceschet & Costantini, 2011), or the Italian National Scientific Qualification (Bologna et al., 2021a). Others have investigated the assessments of departments (Aksnes, 2003), research groups (van Raan, 2006) and disciplines (Poggi et al., 2019). Few works have examined this topic at the individual level (Bornmann et al., 2008; Bornmann & Daniel, 2006; Nederhof & Van Raan, 1987), while many have analyzed the correlation between bibliometrics and research performances (Franceschet, 2010; Leydesdorff, 2009). Recent works have also analyzed the correlation between traditional bibliometrics and altmetrics by also taking into account quality assessment procedures performed by peers (Bornmann & Haunschild, 2018; Nuzzolese et al., 2019).

This work is a continuation of our previous study on the growth of open data available in scientific knowledge graphs and its use in research assessment exercises, rooted in the context of Italian National Scientific Qualification (NSQ) (Di Iorio et al., 2019). The NSQ is the national assessment exercise which establishes whether a scholar can apply to professorial academic positions as Associate Professor (AP) and Full Professor (FP). It combines both quantitative and qualitative evaluation methods. First, it makes use of bibliometrics, as candidates have to pass specific bibliometric thresholds. Then, a peer-review process is employed, where candidates'

CVs are assessed by a committee (a.k.a. the commission) to make the final decision. The bibliometrics are calculated on data collected from commercial databases, i.e. Scopus (Baas et al., 2020) and Web of Science (Birkle et al., 2020), just like in other national academic assessments. This is due to the fact that open citation indexes still offer a limited coverage and are still very few (van Eck et al., 2018).

In (Di Iorio et al., 2019), we simulated the first phase of the NSQ for the applicants for the Computer Science discipline, by computing the bibliometrics considered in the evaluation process using only open data from three scientific knowledge graphs: Crossref (https://www.crossref.org/, an official DOI Registration Agency which makes available open bibliographic metadata) (Hendricks et al., 2020), DBLP (https://dblp.uni-trier.de/, a computer science bibliography website), and COCI (http://opencitations.net/index/coci, a citation index of open DOI-to-DOI citations) (Heibi et al., 2019a). Then, we compared the official results computed using data from Scopus and Web of Science, which is not freely available, with those obtained by using the aforementioned open scientific knowledge graphs. This experiment showed that there is still a large difference between open and closed bibliographic metadata and citations, and that open data cannot yet be used as the only source for academic assessment tasks.

In this present work, two years later, we expand our investigation going beyond Computer Science, and we consider all the *citation-based disciplines*. In the NSQ, disciplines are divided into two categories, i.e. *citation-based disciplines* (CDs) and *non-citation-based disciplines* (NDs)[1]. Depending on which category the discipline belongs to, candidates to that discipline are assessed using different bibliometrics in the first part of the process.

This paper aims to present the methods and results of our study and its implications. It is structured as follows. Section "Background" provides the necessary background information, by providing more information about the NSQ process and giving an overview of the current state of open citations. Section "Methods and materials" presents the sources that have been used in this work to collect the necessary metadata and citation data. Section "Results" discusses the results and lessons derived from our analysis before we draw conclusions. Finally, in Section "Discussion e Threats to Validity", we discuss the outcomes of our experiment and point to new possible research directions to this regard.

---

[1] In the NSQ nomenclature, CDs and NDs are actually tagged as *bibliometric disciplines* and *non-bibliometric disciplines*, respectively. However, this terminology is inconsistent with that of the scientometrics community, since, in the first phase of the NSQ, all candidates are assessed using bibliometrics. The two categories will be renamed CDs and NDs in our study, as two citation-based metrics are used to evaluate candidates to the first category, and none is used for the second category.

# Background

Before we describe our experiment, it is necessary to provide readers with background information on the Italian NSQ and the open citations movement.

## The Italian National Scientific Qualification (NSQ)

Italian Law of December 30th 2010 n.240 (L. 240/2010, 2011) instituted the NSQ, a nation-wide research assessment exercise similar to other procedures already established in other countries. The first two terms of the NSQ took place in 2012 and 2013, followed by a session that lasted from 2016 to 2018, with 1 term in 2016, 2 terms in 2017 and 2 terms in 2018.

The NSQ consists of two distinct qualification procedures aimed at attesting that a scholar has reached the necessary scientific maturity to hold the academic positions of Full Professor (FP), or Associate Professor (AP). However, passing the NSQ does not grant a tenure position. Each university is in charge of opening and filling vacant positions in compliance with the local hiring regulations.

Ministerial Decree of June 14th 2012 (D.L. 2012, 2012) defines a taxonomy of 184 Recruitment Fields (RF) organized in groups and divided into 14 different Scientific Areas (SA). SAs are equivalent to vast academic disciplines and RFs correspond to specific scientific fields of study. In the taxonomy, RFs are identified by an alphanumeric code in the form AA/GF. AA is a number indicating the SA, G is a single letter identifying the group, and F is a digit indicating the single RF. For instance, Neurology's code is 06/D5, where 06 indicates the SA Medicine and D indicates the group Specialized Clinical Medicine (D.L. 2012, 2012). When applying to the NSQ, scholars can choose to be considered for more RFs at a time. Since each RF has its own assessment rules, the candidate may pass the qualification in some disciplines and not in others.

For each RF, an evaluation committee (a.k.a. a commission) is appointed by the Italian Ministry of University and Research (MUR). Each commission is composed of five full professors and is in charge of assessing candidates to their RF for both FP and AP roles.
In the first phase of the NSQ, the candidate's academic maturity is evaluated based on three metrics. For each RF and academic position (i.e. FP and AP) ANVUR (https://www.anvur.it/en/), the National Agency in charge of the NSQ, set and officially released three thresholds for the metrics. Candidates are expected to exceed at least two out of the three thresholds to pass the first step of the ASN.

The three metrics used at this stage vary depending on whether the candidate has applied to a CD or a ND. CDs are all RFs in the first nine SAs (01-09) and NDs are the last five SAs (10-14). Exceptions to this rule are 08/C1, 08/D1, 08/E1, 08/E2, 08/F1 that are considered NDs despite being in the first nine SAs, and all the RFs in Psychology (11/E) that are CDs despite belonging to the last five SAs. Candidates applying to CDs are evaluated using the following metrics:

- the number of their journal papers;
- the total number of citations received;
- their h-index.

While, candidates applying to NDs are evaluated using the following metrics:

- number of their journal papers and book chapters;
- number of their papers published on Class A journals[2];
- number of their published books.

Citation-based metrics are not used for the evaluation of candidates to NDs since, according to ANVUR, no sufficiently complete citation database exists for said disciplines. All the thresholds for each RF, as well as each candidate's bibliometrics, are calculated by ANVUR using data collected from Scopus and Web of Science.

Since citation and paper counts increase over time, normalization based on the scientific age of candidates' publications (i.e. the number of years since the first publication) is applied to the computation of most of these metrics. Moreover, for candidates to the role of FP, only the publications that are less than 15 years old are considered and, for candidates to the role of AP, only the publications that are less than 10 years old are considered. After this phase, applicants are evaluated using the detailed CV about their research accomplishments that they are required to submit when applying to the NSQ.

It is worth mentioning that the experiment introduced in this article focuses only on the first phase of the evaluation process. Indeed, we compare the possibility for each candidate to CDs to exceed thresholds when using open data. We do not take into consideration the final subjective evaluation of the commission (i.e. the second step of the NSQ).

Several studies have been conducted on the NSQ by the research community, such as the quantitative analysis of the 2012 NSQ in (Marzolla, 2015, 2016), the study on the NSQ's impact on self-citation rate (Baccini et al., 2019; Peroni et al., 2020; Scarpa et al., 2018), the investigation of the relationship between the outcomes of the NSQ and the actual scientific excellence of the applicants (Abramo & D'Angelo, 2015), the correlation between altmetrics and traditional indicators (Nuzzolese et al., 2019). Our aim is not to evaluate the reliability of the NSQ, nor to examine its influence and consequences, but to investigate whether such an assessment procedure could be performed without making use of commercial citation indexes.

---

[2] The List of Class A journals is periodically released by the National Agency for the Assessment of Universities and Research (in Italian, *Agenzia Nazionale di Valutazione del sistema Universitario e della Ricerca*, or ANVUR), available at https://www.anvur.it/attivita/classificazione-delle-riviste/classificazione-delle-riviste-ai-fini-dellabilitazione-scientifica-nazionale/elenchi-di-riviste-scientifiche-e-di-classe-a/ (Last accessed 13 January 2021).

# The open citations movement

The first project to introduce for the very first time the open availability of open bibliographic and citation data by the use of Semantic Web (Linked Data) technologies was the OpenCitations Corpus, in 2010, as the main output of a project funded by JISC (Shotton, 2013). However, the availability of open citation data recently changed drastically with the introduction of Initiative for Open Citations (I4OC, https://i4oc.org), in April 2017.

The Initiative was born with the idea of promoting the release of open citation data, and explicitly asked the main scholarly publishers, who deposited their citations on Crossref (https://crossref.org), to release them in the public domain. As of 11 March 2021, we have open reference lists of more than 47 million articles' metadata deposited in Crossref by 2,488 distinct publishers. Also, a list of important stakeholders – such as libraries, consortiums, projects, organizations, companies, and, in particular, founders (Shotton, 2018) – are supporting the movement, several international events (e.g. two editions of the Workshop on Open Citations and Open Scholarly Metadata, several editions of the WikiCite conference, and the Workshop on Open Citations: Opportunities and Ongoing Developments at ISSI 2019) have been organised for promoting the open availability of citation data and bibliographic metadata, and several infrastructures, projects and datasets have been released so far so as to leverage the open citation data available online, such as OpenCitations (http://opencitations.net) (Peroni & Shotton, 2020), OpenAIRE (Rettberg & Schmidt, 2012), WikiCite and Scholia (Nielsen et al., 2017), and the Open Ucranian Citation Index (Cheberkus & Nazarovets, 2019).

# Methods and materials

The first step of our analysis consisted in computing the metrics proposed by the NSQ by using only open data for each candidate in the Scientific Areas (SAs) fully related to CDs, namely: Mathematics and Informatics (SA 01), Physics (SA 02), Chemistry (SA 03), Earth Sciences (SA 04), Biology (SA 05), Medicine (SA 06), Agricultural and Veterinary Sciences (SA 07), Industrial and Information Engineering (SA 09)[3]. There are also some exceptions of RFs that are considered as CDs even if they belong to SAs that are also related to NDs. These were also included in our analysis: RF 08/A1 (Hydraulics, Hydrology, Hydraulic And Marine Constructions), RF 08/A2 (Sanitary And Environmental Engineering, Hydrocarbons And Underground Fluids, Safety And Protection Engineering), RF 08/A3 (Infrastructural And Transportation Engineering, Real Estate Appraisal And Investment Valuation), RF 08/A4 (Geomatics), RF 08/B1 (Geotechnics), RF 08/B2 (Structural Mechanics), RF 08/B3 (Structural Engineering), and the four RFs in Psychology, namely RF 11/E1 (General Psychology,

---

[3] Since all the Scientific Area names and the Recruitment Field names have been defined only in Italian, we use here the official English translation provided by the Italian National University Council (CUN), the elected body representing the Italian University System, which is available at https://www.cun.it/documentazione/academic-fields-and-disciplines-list/.

Psychobiology and Psychometrics), RF 11/E2 (Developmental and Educational Psychology), RF 11/E3 (Social Psychology and Work and Organizational Psychology), and RF 11/E4 (Clinical and Dynamic Psychology).

To do so, we first collected the CVs of all applicants to the five sessions of the NSQ 2016-2018, which have been made publicly available on the ANVUR website for a short period of time. We collected 14,084 CVs for FP and 27,599 CVs for AP. Note that each CV corresponds to a single application, and that the same applicant may apply multiple times (i.e. in more than one session) for multiple levels.

The next step consisted in collecting the list of the DOIs of all the publications that each candidate specified in her/his CV, thus excluding all the publications that do not have associated a DOI. This led us to miss some publications, for instance the workshop articles in the CEUR-WS volumes which are published without a DOI (though Scopus takes track of them). However, we expect that the loss in term of citations is rather limited, considering that the most relevant works and their extensions are usually identified by a DOI, as in the case of journal articles and conference proceedings. The extraction was performed as follows: we extracted the text from each CV (which was originally in PDF format), and searched for valid DOIs using a simple pattern matching approach to produce the publication list. We also managed to solve some recurring issues (e.g. encoding problems for DOIs copied and pasted from the web, DOIs split into several lines, etc). The DOI system Proxy Server REST API (http://www.doi.org/factsheets/DOIProxy.html#rest-api) have been used to verify the existence and validity of the collected DOIs, while we have used the Crossref APIs (https://api.crossref.org) to retrieve all the publication types associated with each DOI, so as to understand if a certain DOI was a journal article or another kind of publication.

Some CVs had irregularities in their internal structure and the process could not extract in a reliable way the information we needed for our analysis. Discarding these CVs, we eventually analyzed 13,267 candidates as FP (94.20% of the total) and 26,070 candidates as AP (94,46% of the total).

The collected data have been used to produce a publication list for each candidate's application. In particular, we were able to retrieve 533,335 unique DOIs overall.

All the citations related to the DOIs extracted were gathered from COCI, the OpenCitations Index of Crossref open DOI-to-DOI citations (Heibi et al., 2019a). This scientific knowledge graph is provided by OpenCitations (Peroni & Shotton, 2020), an infrastructure organization dedicated to open scholarship and the publication of open bibliographic and citation data by the use of Semantic Web (Linked Data) technologies. Launched in July 2018, COCI is the first of the Indexes proposed by OpenCitations (http://opencitations.net/index) in which citations are exposed as first-class data entities with accompanying properties, and currently contains 759,516,507 DOI-to-DOI citation links between 60,778,357 distinct bibliographic entities (OpenCitations, 2020).

Considering the COCI dump used for the experiment, the majority of the citations that are not available in COCI comes from just four publishers: Elsevier, the American Chemical Society, IEEE, and University of Chicago Press (Heibi et al., 2019b). This is due to the particular access policy chosen by these publishers when the last dump of COCI was released. It is worth noticing that, recently, Elsevier and the American Chemical Society have decided to release openly the reference lists of their articles on Crossref (https://crossref.org). Thus, future releases of COCI will also include these missing citation data. Of course, the fact that the release of COCI we used in our study did not contain citations from Elsevier's articles could have been a bottleneck to the study we are presenting, since such publisher manages several of the most important journals in all the SAs of the CDs considered in our study, that are valuable sources of citations to other articles.

The source code of the pipeline to collect these data is available as open source at https://github.com/sosgang/asn-open-multidomain, while the data are available on Zenodo and are released with a CC0 waiver (Bologna et al., 2021b).

The core of the experiment consisted in studying the performance of the candidates when using open data instead of closed ones. The following step consists in calculating the three thresholds for each candidate's application against which compare our data. We used the official thresholds proposed by ANVUR directly, even if they were calculated from closed (and, potentially, richer) data. The current values of these thresholds are available in (Bologna et al., 2021b).

**Table 1. Two (real) candidates of the NSQ accompanied by their values for the three metrics used in the NSQ, i.e. number of journal articles, number of citations, and h-index. The number shown refers to those retrieved by means of open data and the real ones calculated in the context of the NSQ. The official NSQ thresholds are shown in the last three columns.**

| id | Open data | | | Official NSQ data | | | Thresholds | | |
|---|---|---|---|---|---|---|---|---|---|
| | #journal articles (A) | #citations (B) | H-index (C) | #journals articles (A) | #citations (B) | H-index (C) | #journals articles (A) | #citations (B) | H-index (C) |
| 1 | 5 | 146 | 4 | 12 | 226 | 6 | 6 | 20 | 3 |
| 2 | 4 | 1 | 1 | 16 | 7 | 2 | 9 | 2 | 1 |

Then, for each metric – i.e. number of journal articles, number of citations received, and h-index – we calculated the percentage of candidates' applications who were able to exceed the thresholds in both our test and the official NSQ. We also measured the amount of applications who exceeded two thresholds over three – thus, passing the first NSQ phase and enabling them to continue the process to get the qualification. It is worth mentioning that we did not compare the values of the metrics directly, as we expected them to be different, rather their contribution to the qualification (i.e. if they were either above or below the thresholds).

For instance, let us consider the two (real) candidates in Table 1. They both applied for the qualification as AP (in two different RFs) and exceeded all three thresholds. The values of their metrics were lower when we only took open data into account in both cases. However, candidate #1 was able to exceed two thresholds anyway. The same did not happen for candidate #2. We investigated these situations to study the relation between open and closed data.

# Results

We measured the percentage of candidates, for all levels and and for all RFs, for which there is agreement between our test and the official NSQ outcome. Table 2 summarizes the data for all 13,267 candidates as Full Professor (FP, Level 1) and 26,070 candidates as Associate Professor (AP, Level 2). The rows detail each metric and the overall agreement, in other words whether the results computed using open data correspond to the official NSQ results in the first phase (i.e. the candidate passes the first NSQ phase if has at least two metrics over three above the thresholds, while he/she fails the qualification otherwise).

**Table 2. The percentage of candidates as FP and AP who achieved the same result in our open data simulation and the official ASN, for each metric.**

|  | Full Professor (13,267 candidates) | Associate Professor (26,070 candidates) |
|---|---|---|
| Overall agreement | 51.07% | 59.83% |
| Journals (A) | 47.58% | 65.70% |
| Citations (B) | 53.80% | 61.36% |
| H-index (C) | 54.93% | 65.01% |

Overall, the results on open data are not yet comparable to those on closed ones. In fact, the agreement is around 50% for FP and around 60% for AP. The fact that the results are worse in the first case than the second is not surprising. Indeed, the thresholds for FP are higher than those for AP and it is more difficult to reach them if the data are limited and incomplete. The agreement is in fact lower considering only the candidates who succeeded in the NSQ.

The contribution of each metric is substantially comparable, and slightly higher on metric C (h-index). It is interesting to note that the metric A (articles in journals) shows the lowest agreement for FP and the highest for APs. This could be affected by two factors: the distribution of applications across the two levels (AP vs. FP), and the lack of a correct and complete classification of the journal articles as discussed in Section "Discussion e Threats to Validity".

While the results are quite aligned across levels, we witness a lot of differences across RFs and SAs. Tables 3 and 4 show the percentages of agreement for each SA and for the two levels. Space limits prevent us from reporting the details of each single RF here but all data are available in the material online. The two tables also show the number of candidates who applied to the RFs related to CDs in each SA, so as to weigh the results.

**Table 3. The percentage of candidates as FP who achieved the same result in our open data simulation and the official NSQ, for each metric and for each SA. The second row also shows the number of candidates who applied for that SA. The 'overall agreement' indicates the percentage of candidates for whom two metrics over three are above or below the thresholds, in both cases (open and closed data).**

| | *SA 01* | *SA 02* | *SA 03* | *SA 04* | *SA 05* | *SA 06* | *SA 07* | *SA 08* | *SA 09* | *SA 11* |
|---|---|---|---|---|---|---|---|---|---|---|
| Number of candidates | 1289 | 1266 | 1234 | 453 | 2235 | 3074 | 854 | 517 | 1988 | 357 |
| Overall agreement | 63.36% | 64.69% | 27.09% | 55.79% | 47.89% | 42.89% | 53.01% | 70.65% | 52.56% | 71.02% |
| Journals (A) | 66.13% | 28.24% | 23.23% | 65.78% | 53.27% | 30.90% | 62.91% | 83.64% | 41.22% | 80.98% |
| Citations (B) | 61.13% | 72.65% | 33.72% | 51.94% | 51.04% | 48.59% | 53.88% | 65.31% | 55.16% | 72.86% |
| H-index (C) | 64.48% | 69.87% | 31.76% | 57.24% | 50.3%8 | 47.45% | 55.28% | 72.77% | 59.40% | 72.67% |

The overall agreement for FP is quite heterogeneous, spanning form ~27% (Area 03) to ~71% (Area 11).

The heterogeneity across SAs is also confirmed when considering each metric separately. The metric A, for instance, goes from ~23% up to ~81%. The same for metric B (from ~33% to ~73%) and metric C (from ~31% to ~72%). However there is a lot of uniformity between the metrics on the same SA, with a few exceptions. SA 02 shows an agreement of ~28% for metric A and ~73% for metric B (thus, a difference of 44%). SA 09 scores ~41% for metric A and ~59% for metric C (18% of difference).

In general, there is no area where the metrics are remarkably high or low. The values higher than 70% are highlighted with a green background in Table 3, while those lower than 30% are in grey. With the exceptions of SAs 08 and 11, whose results are very good, the majority of values range between 30% and 70%. It is worth mentioning also the very low agreement for SA 03.

As shown in Table 4, the results for AP are quite aligned to those obtained for FP even if, in general, the agreement is much higher. In fact, it spans from ~39% (SA 03) to ~77% (SA 02). It is worth noting that the SAs with the lowest and highest agreement are the same as the FP case, respectively SA 03 and SA 11.

The heterogeneity of the metrics is much less pronounced in Table 4, in comparison to Table 3. The metric A goes from ~54% up to ~88%, while B goes from ~47% to ~77% and C from ~44% to ~80%. Again, these highest values are in SA 08 and SA 11.

The uniformity of the metrics within the same SA, on the other side, is more evident. In the worst cases, in fact, there is a difference of 15% between metric A and B for SA 02.

Note also that there is no SA where the agreement is lower than 30% (gray background) and, compared to Table 3, much more values over 70% (green background) or close to that (for instance the metric A for SA 07 and the metric C for SA 01).

**Table 4. The percentage of candidates as AP who achieved the same result in our open data simulation and the official NSQ, for each metric and for each SA. The second row also shows the number of candidates who applied for that SA. The 'overall agreement' indicates the percentage of candidates for whom two metrics over three are above or below the thresholds, in both cases (open and closed data).**

|  | SA 01 | SA 02 | SA 03 | SA 04 | SA 05 | SA 06 | SA 07 | SA 08 | SA 09 | SA 11 |
|---|---|---|---|---|---|---|---|---|---|---|
| Number of candidates | 1789 | 2028 | 2030 | 721 | 6510 | 6497 | 1461 | 844 | 3375 | 815 |
| Overall agreement | 63.98 % | 73.84 % | 38.85 % | 55.97 % | 56.25 % | 58.89 % | 62.77 % | 71.50 % | 57.57 % | 77.76 % |
| Journals (A) | 72.42 % | 55.33 % | 54.29 % | 70.24 % | 68.29 % | 57.71 % | 69.38 % | 88.19 % | 65.84 % | 78.88 % |
| Citations (B) | 65.24 % | 77.70 % | 47.25 % | 55.74 % | 58.51 % | 60.10 % | 62.55 % | 64.52 % | 60.63 % | 76.90 % |
| H-index (C) | 68.57 % | 79.35 % | 44.44 % | 60.74 % | 61.20 % | 65.16 % | 70.79 % | 72.24 % | 61.45 % | 80.78 % |

In order to study the relation between the two levels, Figure 1 shows the overall agreement for all the SAs in increasing order, for FP and AP. The two trends are comparable, with a homogeneous improvement for the second group.

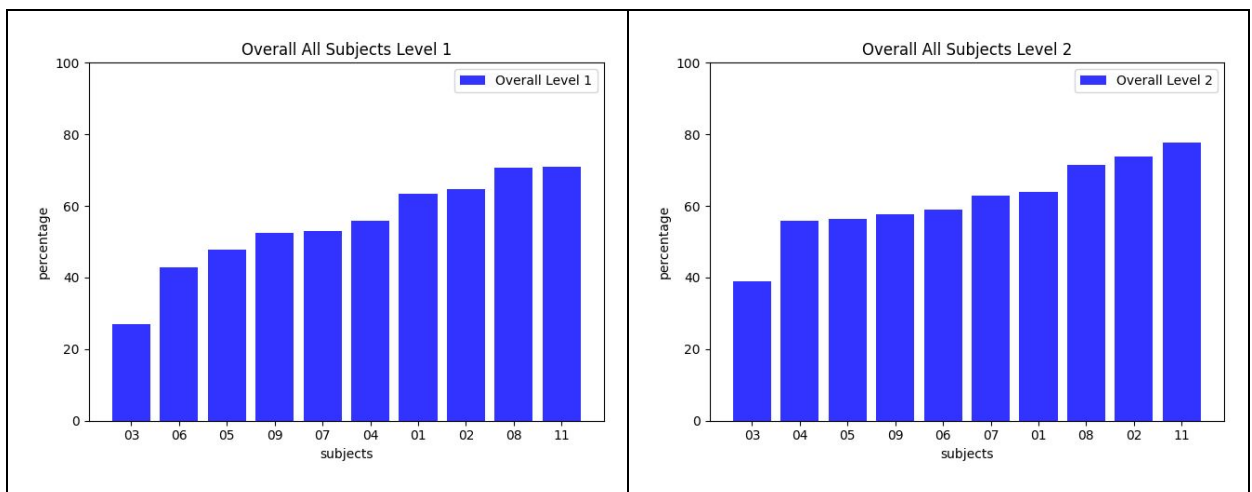

**Figure 1. The overall percentage of agreement with the official NSQ results for all the SAs, in increasing order, for either level 1 (on the left) and level 2 (on the right).**

Note also that the relative order between the SAs is different in the two cases, even if the areas with the minimum and maximum agreement (respectively SA 03 and SA 11) do not change.

Looking at the details of each RF we found some peculiar situations worth highlighting. Figure 2 shows the agreement on the RFs with the worst performance, 03/D2 (Drug Technology, Socioecodomics and Regulations) and 09/D3 (Chemical Plants and Technologies). The amount of open data is extremely limited in these fields and clearly not yet ready to be used for evaluation purposes.

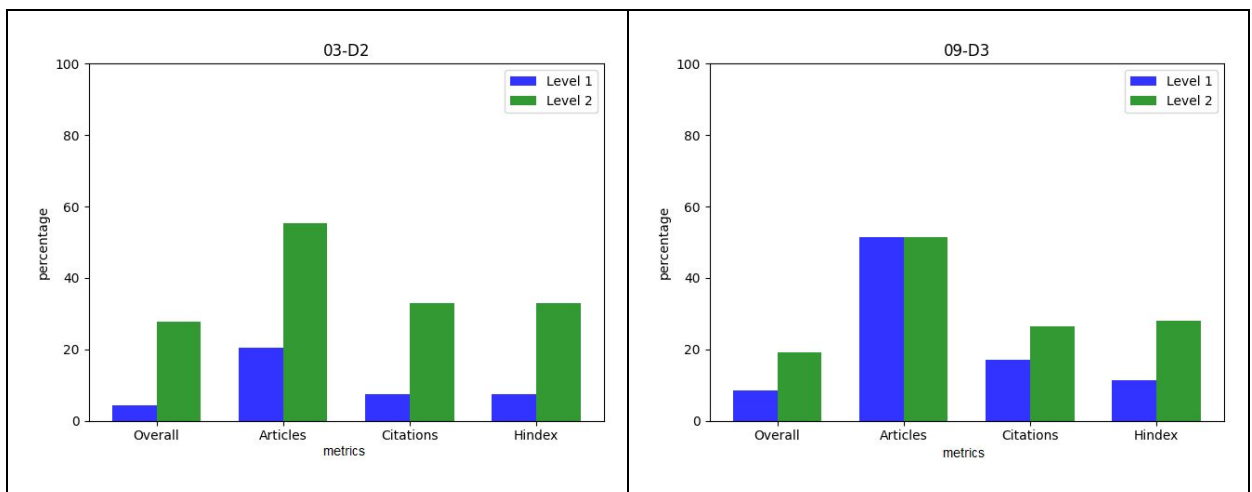

**Figure 2. The worst results, achieved on RF 03/D2 and RF 09/D3, with low agreement for all indicators, and in particular for level 1 (FP). The Y-axis reports the percentage of agreement with the official NSQ results.**

The opposite scenario happens for RF 08/A3 (Infrastructural And Transportation Engineering, Real Estate Appraisal And Investment Valuation) and 08/A4 (Geomatics), as shown in Figure 3, which performed well for all the metrics. These results are comparable to those of the official NSQ evaluation, thus open data could be taken into account in the near future. Similar results were also reported for 11/E2 (Developmental and Educational Psychology) and 11/E4 (Clinical and Dynamic Psychology).

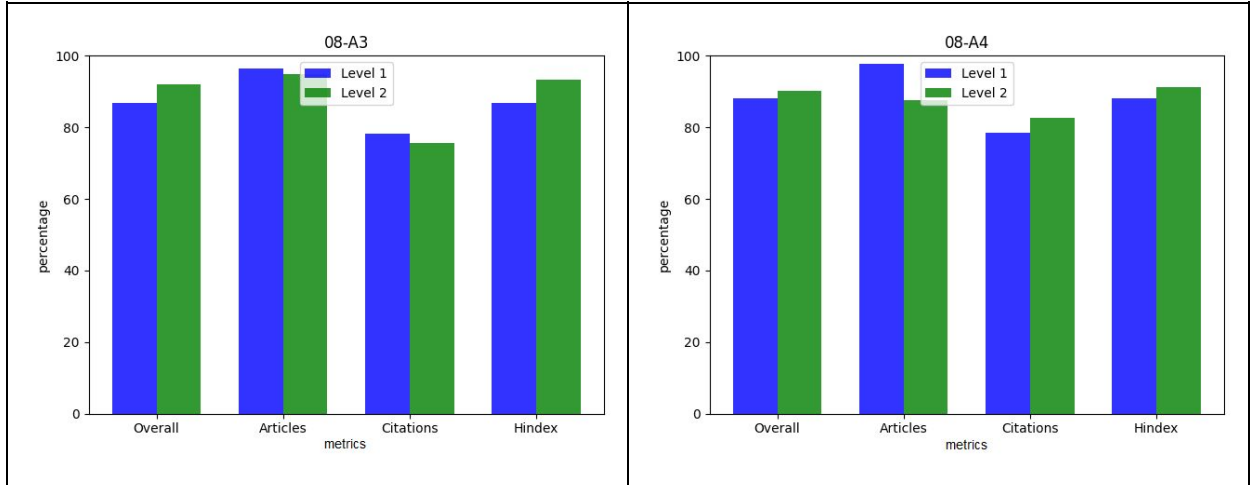

**Figure 3. The best results, achieved on RF 08-A3 and 08-A4, with very high agreement for all metrics. The Y-axis reports the percentage of agreement with the official NSQ results.**

Figure 4 shows a trend we found in many RFs, among which RF 02/A1(Experimental Physics of Fundamental Interactions) and RF 02/C1 (Astronomy, Astrophysics, Earth and Planetary Physics), as well as 09/E1 (Electrical Engineering) and 02/A2 (Theoretical Physics of Fundamental Interactions) not shown here. The metrics B and C go very well but there is limited agreement on metric A (i.e. number of journals). This happens because Crossref contains partial data about the journals in the field.

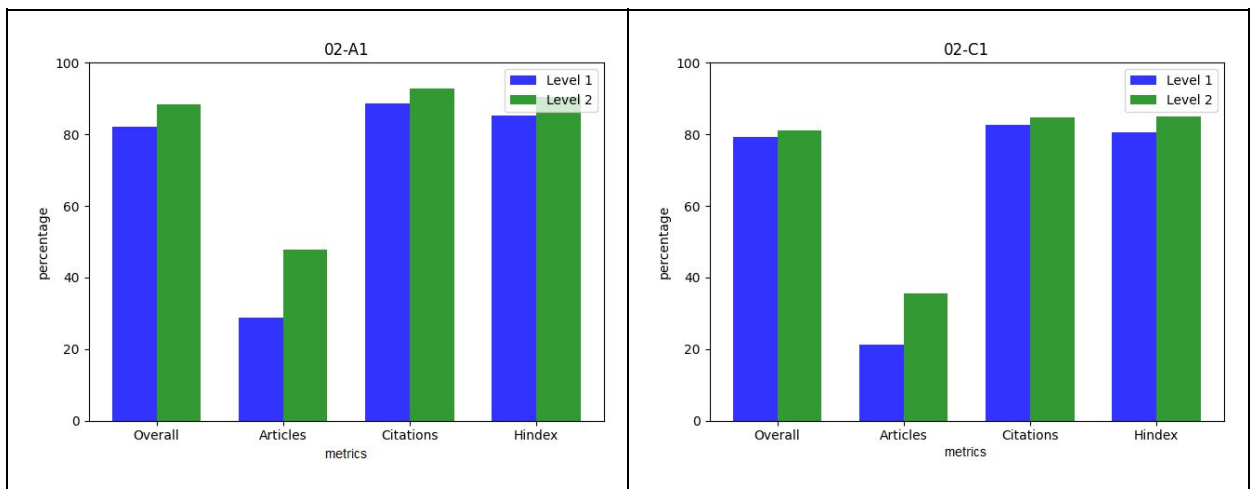

**Figure 4. The results on RF 02/A1 and RF 02/C1. The agreement is high for all the metrics, with the exceptions of the number of journal articles.**

Finally, Figure 5 shows the most common situation: the agreement is not so high but uniform for all metrics and higher for AP. This happens in 09/G2 (Bioengineering) and 09/D2 (Systems, Methods and Technologies of Chemical and Process Engineering), but also in 09/G1 (Systems and Control Engineering), 07/I1 (Agricultural Microbiology) and others not shown here.

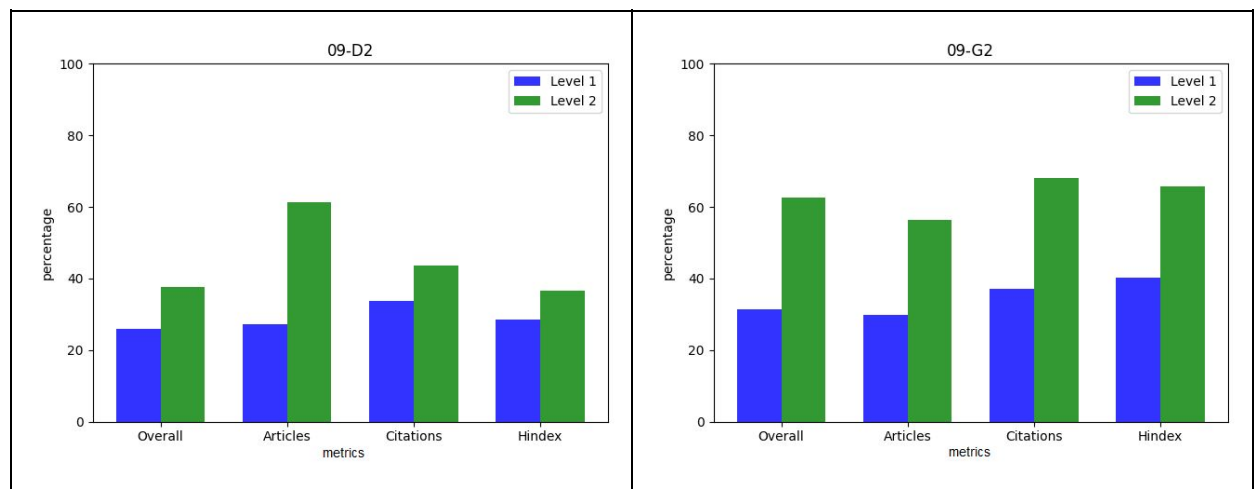

**Figure 5. The results on RF 09/D2 and 09/G2, common with many other RFs: the agreement is quite low, and much higher for level 2 (AP) than level 1 (FP). The Y-axis reports the percentage of agreement with the official NSQ results.**

The other RFs did not show regularities that can be generalized. Metrics perform in different ways, even within the same RF, depending on the amount and quality of open data available for that domain. The full set of diagrams is available in (Bologna et al., 2021c).

# Discussion e Threats to Validity

Our results confirmed what, two years ago, we had studied for the Computer Science domain in (Di Iorio et al., 2019), with some important differences.

First of all, we concluded that **open citation data available in OpenCitations' COCI are not yet complete to substitute the closed data** used by ANVUR within the NSQ. There are some exceptions, for instance in SA 08 and SA 11, where the amount and quality of open data is very high.

The overall trend is positive anyway and we believe that other SAs will be ready to adopt open data for evaluation in the future. It is also interesting the case of Computer Science (RF 01/B1) in which the agreement of metrics B and C has increased in comparison to our previous work. This is what we expected considering the large amount of data that has been added to the latest version of COCI compared to that used in (Di Iorio et al., 2019).

The behavior on the metric A (number of journal articles) was instead different from our previous analysis, since it presented a lot of irregularities and in many cases was lower than others. The classification of the publications proved not to be equally good for all RFs. Such inaccuracy contributed to the low level of agreement of this metric, even when others performed well.

This leads us to another conclusion: **domain-specific open scientific knowledge graphs should also be considered for the evaluation**, in order to get better results. In our previous experiment, in fact, we witnessed a remarkable improvement of the agreement when we also considered data from DBLP, which is a repository of metadata about Computer Science publications. This source is indeed very accurate for computer scientists since almost all their publications are listed there. Other domains do not have a DBLP counterpart, or they have similar sources that we did not take into account at this stage. In the future we also plan to extend our analysis to consider more accurate data to improve the labelling of candidates' publications.

A further conclusion that we have drawn from our experiments is that, currently, the **open data available in the two scientific knowledge graphs used are more suitable for AP than for FP** evaluation. We speculate that this is due to the higher values of the thresholds used for the evaluation: it is more difficult to reach these thresholds if the data are taken from limited sources.

Some threats to validity are worth highlighting about our work. First of all, as for our previous paper, we used the NSQ official thresholds on open data too. This is not optimal but we had no undisputable algorithm to re-calculate thresholds on our dataset in a consistent way with the NSQ procedure, since details about that were not published by ANVUR. While in (Di Iorio et al., 2019) we experimented with some tuning of the Computer Science thresholds, we did not repeat the same analysis here. This is planned for future extensions of this work, together with the integration of machine-learning techniques to perform the analysis.

Another limitation of our analysis is related to the input data. We only considered the data automatically extracted from the candidates CVs but these might be incomplete and inaccurate. More data could be collected by also considering the publications of the candidates available on other sources. Since COCI does not contain information about authors, we plan to extend our work by using other data sources that have this kind of information like MAG or Crossref. Alignment and integration mechanisms could increase the input base and, thus, generate higher agreement.

Finally, it is worth remarking a bias introduced in the analysis by the 2,152 candidates for whom we were not able to collect any DOI. These were included in the analysis but generated a subtle phenomenon. The case of a candidate with no DOIs that did not pass the thresholds in the official NSQ was considered an agreement. In fact, she/he was under the thresholds for both closed and open data. This is disputable due to the total lack of data. On the other hand, we decided not to exclude these candidates since the empty list of DOI is exactly the input that would be available to ANVUR if they had used open data only.